\documentclass[a4paper,12pt]{amsart}
\usepackage{amsmath,amsfonts,amssymb}

\def\C{\mathbb{C}}
\def\N{\mathbb{N}}
\def\R{\mathbb{R}}
\def\Z{\mathbb{Z}}
\def\g{\mathfrak{g}}

\def\d{\delta}

\def\d0{\frac{d}{d \tau}\Big|_{\tau=0}}

\begin{document}
\title[Gauge transformations and integrable systems]{Gauge transformations and symmetries of integrable systems}

\author[T. Fukuyama, K. Kamimura, S. Kre\v{s}i\'{c}-Juri\'{c}, S. Meljanac]
{Takeshi Fukuyama, Kiyoshi Kamimura, Sa\v{s}a Kre\v{s}i\'{c}-Juri\'{c}, Stjepan Meljanac}

\address[T. Fukuyama]{Department of Physics, Ritsumeikan University, Shiga, Kusatsu 525-8577, Japan}
\address[K. Kamimura]{Department of Physics, Toho University, Funabashi 247-8510, Japan}
\address[S. Kre\v{s}i\'{c}-Juri\'{c}]{Faculty of Natural and Mathematical Sciences, University of Split,
Teslina 12, 21000 Split, Croatia}
\email{skresic@fesb.hr}
\address[S. Meljanac]{Rudjer Bo\v{s}kovi\'{c} Institute, Bijeni\v{c}ka cesta b.b., 10000 Zagreb, Croatia}

\begin{abstract}
We analyze several integrable systems in zero-curvature form within the framework of $SL(2,\R)$
invariant gauge theory. In the Drinfeld-Sokolov gauge we derive a two-parameter family of nonlinear
evolution equations which as special cases include the Kortweg-de Vries (KdV) and Harry Dym
equations. We find residual gauge transformations which lead to infinitesimal symmetries of this
family of equations. For KdV and Harry Dym equations we find an infinite hierarchy of such symmetry
transformations, and we investigate their relation with local conservation laws, constants of the
motion and the bi-Hamiltonian structure of the equations. Applying successive gauge transformations
of Miura type we obtain a sequence of gauge equivalent integrable systems, among them the modified
KdV and Calogero KdV equations.
\end{abstract}


\maketitle

\section{Introduction}

Integrable systems in $1+1$ dimensions have been discussed extensively so far, and various
techniques to analyze their integrability have been developed using the Lax or zero-curvature
representation, the inverse scattering method, the Hirota's bilinear method, the Painlev\'e
property, etc. \cite{zakharov2}. Integrable systems have also been investigated within the
framework of factorization problems on loop groups and infinite dimensional Grassmannian spaces
\cite{segal}. Gauge theories play a particular role in the investigation of integrable systems. In
$1+1$ and higher dimensions there is a well-known conjecture by Ward \cite{ward} that any
integrable systems are derived from (anti) self-dual Yang-Mills equations.

In this paper we examine several integrable systems from the point of view of $SL(2,\R)$ invariant
gauge theory. We consider infinitesimal gauge transformations acting on $sl(2,\R)$-valued fields
which satisfy the zero-curvature equation. We analyze the transformations which leave the given
gauge fixing conditions invariant (i.e. residual transformations), and thus generate infinitesimal
symmetries of evolution equations. Recall that if $u(x,t)$ is a solution of the evolution equation
$u_t=K[u]$ where $K[u]$ depends on $x$, $u$, and a finite number of $x$-derivatives of $u$, then an
infinitesimal symmetry is a transformation $u(x,t)\mapsto \widetilde u(x,t)= u(x,t)+\tau y(x,t)$
such that the deformed function $\widetilde u(x,t)$ is also a solution of the given equation to
first order in $\tau$. At least formally the infinitesimal transformations allow one to generate
new solutions from old ones. Symmetries of evolution equations have also been studied by prolonged
group actions on jet spaces, as discussed in Olver \cite{olver}. In this approach a local group of
transformations acts on the space of independent and dependent variables, whereas gauge
transformations considered here are globally defined groups acting on the space of dependent
variables. For a particular class of evolution equations we investigate the relationship between
residual gauge transformations, conservation laws and bi-Hamiltonian structure. We show that the
conservation laws and the bi-Hamiltonian structure of the Korteweg-de Vries (KdV) and Harry Dym
(HD) equations are directly related to an infinite hierarchy of residual transformations, and we
thus provide a new interpretation of some previously known results from a unified perspective. We
also investigate the Miura transformations from the gauge theoretical point of view. Since we do
not incorporate the spectral parameter, i.e. we do not consider the loop algebra $sl(2,\C)\otimes
\C[\lambda,\lambda^{-1}]$, the class of evolution equations obtained here is different from that
considered in \cite{drienfeld}.

The outline of the paper is as follows. In Sec. 2 we consider the Drinfeld-Sokolov gauge on the Lie
algebra $sl(2,\R)$ and we derive a two-parameter family of nonlinear evolution equations which
include as special cases the well-known KdV and HD equations. We find residual gauge
transformations which lead to infinitesimal symmetries of this family of equations. A different
derivation of KdV as a reduction of the self-dual Yang-Mills equations on $\R^4$ was given in
\cite{mason} (see also \cite{ablowitz}). In Sec. 3 we examine more closely the residual
transformations for the KdV equation. We prove the existence of an infinite hierarchy of such
transformations, and we show that each transformation generates a local conservation law. We
further show that the residual transformations are related to an infinite hierarchy of first
integrals found in \cite{wadati} and the Lenard recursion operator for the bi-Hamiltonian structure
of KdV. Furthermore, by applying successively the Miura-type gauge transformations to the KdV
equation we obtain a sequence of gauge equivalent integrable systems, among them the modified KdV
(mKdV) and Calogero KdV (CKdV) equations. By using residual transformations we find nonlocal
symmetries of the mKdV equation. Similar considerations are applied to the Harry Dym equation in
Sec. 4. Finally, in Sec. 5 we note that by changing the gauge fixing condition and generalizing the
gauge group to $SL(2,\C)$ on obtains other classical integrable systems such as the sine-Gordon and
nonlinear Schr\"{o}dinger equation.

\section{General framework}
\label{general_framework}

Let $G$ be a matrix Lie group with Lie algebra $\g$. Consider an overdetermined linear system of
equations for a wave function $\psi$,
\begin{equation}\label{system1}
\frac{\partial \psi}{\partial x_k} = A_k \psi, \quad 1\leq k\leq m,
\end{equation}
where $A_k (x)\in \g$ is an $n\times n$ matrix valued gauge field and $\psi (x)$ is a vector valued function of $x\in \R^m$.
The consistency condition $\partial_k \partial_l \psi = \partial_l \partial_k \psi$ requires that the gauge fields satisfy the
zero-curvature condition \cite{zakharov}
\begin{equation}\label{ZCC}
\frac{\partial A_k}{\partial x_l} - \frac{\partial A_l}{\partial x_k} +[A_k,A_l]=0, \quad 1\leq k,l\leq m.
\end{equation}
Equation \eqref{ZCC} represents a hierarchy of nonlinear partial differential equations (PDEs) for components of the gauge fields.
Equations which can be represented in this form are often said to be integrable, although the proof of integrability
in each particular case is a nontrivial problem. The zero-curvature representation is used as a basis for application
of geometric and algebraic techniques in the study of nonlinear PDEs, and it is also used in classification of integrable
systems by Lie algebraic methods (see e.g. \cite{fadeev} and \cite{babelon}). The Lie algebras considered here are finite
dimensional, although a number of integrable systems can be written in zero-curvature form on infinite dimensional loop algebras.
In this approach the equation at hand is integrated by solving a Riemann-Hilbert type factorization problem on the
underlying loop group (see e.g. \cite{dorfmeister}).

The system of equations \eqref{system1}-\eqref{ZCC} is invariant under the gauge transformation
\begin{align}
\psi &\mapsto g\psi,  \\
A_k &\mapsto \Gamma_g A_k \equiv g A_k g^{-1} + \frac{\partial g}{\partial x_k} g^{-1}, \label{gamma}
\end{align}
where $g\in G$. In general, this transformation changes the particular form of the equations
represented by \eqref{ZCC}. Two systems of equations are said to be gauge equivalent if their
zero-curvature representations are related by a gauge transformation. For examples of gauge
equivalent systems see \cite{fadeev}. Gauge transformations can be used to map solutions of one
system into the other \cite{kresic}. In this paper we shall consider solutions of the system
\eqref{system1}-\eqref{ZCC} which are gauge equivalent. Such solutions have the pure gauge form
\begin{equation}
\psi = g\psi_0, \quad A_k = \frac{\partial g}{\partial x_k} g^{-1},
\end{equation}
for some constant vector $\psi_0\in \C^n$ and $g\in G$. We note that, as pointed out by Wu and Yang \cite{wu}, in some
cases there exist gauge fields which give rise to the same field strength but are not related by any gauge
transformations.

The zero-curvature condition \eqref{ZCC} gives rise to integrable systems of a fairly general
nature. However, one is usually interested in systems involving a smaller number of functions. This
is achieved by imposing gauge fixing conditions which lead to a reduced set of differential
equations for independent field components. In the following we discuss integrable systems in $1+1$
space-time dimensions for $sl(2,\R)$-valued fields. By parameterizing the gauge fields explicitly
the linear problem becomes
\begin{equation}
\frac{\partial \psi}{\partial x} = A_1 \psi, \quad \frac{\partial \psi}{\partial t} = A_2\psi.
\end{equation}
where
\begin{equation}\label{fields}
A_1=\begin{pmatrix} R & S \\ -T & -R\end{pmatrix}, \quad A_2=\begin{pmatrix} p & u \\ q & -p\end{pmatrix}.
\end{equation}
Here $\psi=(\psi_1\, \psi_2)^t$, and the minus sign for $T$ is introduces for convenience. The
zero-curvature condition \eqref{ZCC} yields
\begin{align}
R_t-p_x &= -qS-uT, \label{ZCC1} \\
S_t-u_x &= 2pS-2uR, \label{ZCC2} \\
T_t+q_x &= -2qR-2pT, \label{ZCC3}
\end{align}
where the subscripts denote partial derivatives. There are three gauge freedoms for $A_1$ and $A_2$ since the six field components
are related by three differential equations. First, we fix two of them by requiring
\begin{equation}\label{DS}
R=0 \quad \mbox{and}\quad S=1.
\end{equation}
This is called the Drinfeld-Sokolov (DS) gauge \cite{drienfeld}. In this gauge the fields are given by
\begin{equation}\label{DSfields}
A_1^{(DS)}=\begin{pmatrix} 0 & 1 \\ -T & 0\end{pmatrix}, \quad A_2^{(DS)}=\begin{pmatrix} -\frac{1}{2}u_x & u \\ -\frac{1}{2}u_{xx}-uT &
\frac{1}{2}u_x \end{pmatrix},
\end{equation}
and the zero-curvature equation for $A_1^{(DS)}$ and $A_2^{(DS)}$ is equivalent to
\begin{equation}\label{ZCCTu}
T_t = 2u_x T+uT_x+\frac{1}{2}u_{xxx}.
\end{equation}

We are interested in gauge transformations which preserve the gauge fixing condition \eqref{DS},
and thus leave equation \eqref{ZCCTu} invariant. The existence of such residual transformations
leads to symmetries of evolution equations and to an infinite hierarchy of conservation laws.

Define the map $K\colon sl(2,\R)\to \R^2$ by $K\left[\left(\begin{smallmatrix} a & b \\ c & -a
\end{smallmatrix}\right)\right]=(a,b-1)$, and consider the subvariety $\mathcal{S}_K = \{A\in
sl(2,\R)\mid K(A)=0\}$ which consists precisely of the matrices in the DS gauge. We note that $K$
is of maximal rank since the Jacobian of $K$ is
\begin{equation}
J_K = \begin{pmatrix} 1 & 0 & 0 \\ 0 & 1 & 0 \end{pmatrix}.
\end{equation}
Fix $L \in sl(2,\R)$ and consider the action of the one-parameter gauge group $G=\{\exp(\tau L)\mid
\tau \in \R\}$ on $sl(2,\R)$ by the gauge transformation \eqref{gamma}. Since $K$ is of maximal
rank, $G$ is a symmetry group of the set $\mathcal{S}_K\subset sl(2,A)$ if and only if each
component of $K$ satisfies (see e.g. \cite{olver})
\begin{equation}\label{condition1}
\d0 K_i \left(\Gamma_{\exp(\tau L)} A\right)=0 \quad \mbox{for all} \quad A\in \mathcal{S}_K, \quad i=1,2.
\end{equation}
If $L$ satisfies condition \eqref{condition1}, then the gauge transformation $A_1^{(DS)}\mapsto
\Gamma_g A_1^{(DS)}$ preserves the DS gauge and hence equation \eqref{ZCCTu}. In order to find $L$
let $\pi \colon sl(2,\R)\to \R^2$ denote the projection $\pi \left[\left(\begin{smallmatrix} a & b
\\ c & -a \end{smallmatrix} \right)\right]=(a,b)$. Then
\begin{equation}
\d0 K\left(\Gamma_{\exp(\tau L)} A\right) = \pi \left(\d0 \Gamma_{\exp(\tau L)} A\right),
\end{equation}
where the expression in the parentheses defines the infinitesimal gauge transformation by $L$ given by
\begin{equation}
\d0 \Gamma_{\exp(\tau L)} A = [L,A]+\frac{\partial L}{\partial x}.
\end{equation}
Hence, in view of the condition \eqref{condition1} the DS gauge is preserved if and only if for
every matrix $A$ in the DS gauge, $A\in \mathcal{S}_K$, we have
\begin{equation}\label{condition2}
\pi \left([L,A]+\frac{\partial L}{\partial x}\right)=0.
\end{equation}
Setting $A=A_1^{(DS)}$ and evaluating condition \eqref{condition2} we see that $L$ can be
parameterized by an arbitrary function $y (x,t)$ as
\begin{equation}
L=\begin{pmatrix} -\frac{1}{2}y_x & y \\ -\frac{1}{2}y_{xx}-y T & \frac{1}{2}y_x \end{pmatrix}.
\end{equation}
The gauge transformation of $A_1^{(DS)}$ is now given by
\begin{align}
\Gamma_{\exp(\tau L)}A_1^{(DS)} &= A_1^{(DS)}+\tau \left([L,A_1^{(DS)}]+\frac{\partial L}{\partial x}\right)+o(\tau^2) \notag  \\
&= \begin{pmatrix} 0 & 1 \\ -T & 0 \end{pmatrix}+
\tau\begin{pmatrix} 0 & 0 \\ -\frac{1}{2}y_{xxx}-2y_xT-yT_x & 0 \end{pmatrix}+o(\tau^2), \label{gaugeA1DS}
\end{align}
hence the infinitesimal transformation of $T$ is
\begin{equation}\label{Ttransform}
\widetilde T = T+\tau \left(\frac{1}{2}y_{xxx}+2y_xT+yT_x\right).
\end{equation}
Similarly, the gauge transformation of $A_2^{(DS)}$ yields
\begin{align}
&\Gamma_{\exp(\tau L)} A_2{(DS)} = A_2^{(DS)}+\tau \left([L,A_2^{(DS)}]+\frac{\partial L}{\partial t}\right)+o(\tau^2) \notag \\
&= \begin{pmatrix} -\frac{1}{2}u_x & u \\ -\frac{1}{2}u_{xx}-uT & \frac{1}{2}u_x \end{pmatrix} \label{gaugeA2DS} \\
&+\tau \begin{pmatrix} \frac{1}{2} u_{xx}y+\frac{1}{2}u y_{xx}-\frac{1}{2}y_{xt} & y_t-y_x u+y u_x \notag \\
\frac{1}{2} u_x y_{xx}-\frac{1}{2} u_{xx} y_x + (y u_x-y_x u)T-\frac{1}{2} y_{xxt}-(yT)_t &
-\frac{1}{2} u_{xx}y-\frac{1}{2}u y_{xx}+\frac{1}{2}y_{xt} \end{pmatrix} \\
&+o(\tau^2). \notag
\end{align}
Thus, the infinitesimal transformation of $u$ is given by
\begin{equation}\label{utransform}
\widetilde u = u+\tau (y_t-y_x u + y u_x).
\end{equation}
For convenience we denote the infinitesimal deformations by $\delta T =
\frac{1}{2}y_{xxx}+2y_xT+yT_x$ and $\delta u =y_t-y_xu+yu_x$. Since the gauge transformations of
$A_1^{(DS)}$ and $A_2^{(DS)}$ satisfy the zero-curvature equation (which can be also checked
directly to first order in $\tau$ using equations \eqref{gaugeA1DS} and \eqref{gaugeA2DS}), the
functions $\widetilde T = T+\tau \delta T$ and $\widetilde u = u +\tau \delta u$ satisfy equation
\eqref{ZCCTu} to first order in $\tau$. Hence, the maps $T\mapsto \widetilde T$ and $u\mapsto
\widetilde u$ are infinitesimal symmetries of \eqref{ZCCTu}.

The remaining gauge freedom is fixed by imposing a relation between $T$ and $u$. We set
\begin{equation}\label{condition3}
T=\frac{u^{\alpha}}{s}, \quad \alpha\in \mathbb{Z}\setminus \{0\},\; s>0.
\end{equation}
The zero-curvature condition \eqref{ZCCTu} then becomes
\begin{equation}\label{GKdV}
u_t =\frac{\alpha+2}{\alpha}\, u\, u_x + \frac{s}{2\alpha}\, u^{1-\alpha}\, u_{xxx}.
\end{equation}
The motivation for considering the gauge \eqref{condition3} is that \eqref{GKdV} is a two-parameter family of evolution
equations which, as special cases, includes some well-known integrable systems: the KdV equation for $\alpha=1$, $s=2$, and Harry Dym
equation for $\alpha=-2$, $s=1$.

The infinitesimal symmetry of \eqref{GKdV} is easily found by requiring that the function $y$ is
chosen so that $\widetilde T$ and $\widetilde u$ satisfy condition \eqref{condition3} to first
order in $\tau$. This holds if and only if
\begin{equation}\label{dTdu}
\delta T = \frac{\alpha}{s}\, u^{\alpha-1}\, \delta u.
\end{equation}
By substituting \eqref{Ttransform} and \eqref{utransform} into \eqref{dTdu} we find that $y$ is constrained by the
equation
\begin{equation}\label{symmetry}
y_t = \frac{\alpha+2}{\alpha}\, u\, y_x+\frac{s}{2\alpha}\, u^{1-\alpha}\, y_{xxx}.
\end{equation}
In this case the infinitesimal transformation for $u$ is given by
\begin{equation}\label{du2}
\widetilde u = u+\tau \left(\frac{2}{\alpha}\,u\, y_x+ u_x\,y +\frac{s}{2\alpha}\, u^{1-\alpha}\, y_{xxx}\right).
\end{equation}
Any smooth solution of equation \eqref{symmetry} generates a residual gauge transformation of
\eqref{GKdV}.

We noted earlier that equation \eqref{GKdV} is integrable for $\alpha=1,-2$. Then equation
\eqref{symmetry} has infinitely many solutions which generate a hierarchy of conservation laws for
\eqref{GKdV}, as discussed later. Furthermore, a common property shared by all equations
\eqref{GKdV} is the existence of travelling wave solutions. We conjecture that equation
\eqref{GKdV} is integrable for all $\alpha \in \Z\setminus \{0\}$, which is certainly true if
\eqref{symmetry} has an infinite number of solutions for all such $\alpha$. For a discussion of
integrability of \eqref{GKdV} using the Painlev\'e property see \cite{fukuyama}. It is also noted
that for $\alpha=1$ the matrices $A_1^{(DS)}$ and $A_2^{(DS)}$ can be extended to matrix
polynomials in the loop algebra $sl(2,\C)\otimes \C[\lambda,\lambda^{-1}]$ appearing in the
Riemann-Hilbert factorization for the KdV equation (see \cite{dorfmeister}). It would be a very
interesting problem to determine if this construction carries over for all $\alpha$ and if
\eqref{GKdV} can be integrated by solving a Riemann-Hilbert factorization problem.

As stated earlier, equations \eqref{GKdV} admit travelling wave solutions $u(x+vt)$. They are given
explicitly for $\alpha = \pm 1, \pm 2$, and in implicit form for other values of $\alpha$. The
solutions are assumed to satisfy the boundary conditions $u,u^\prime, u^{\prime\prime}\to 0$ as
$|x+vt|\to \infty$. For $\alpha=2k-1$, $k\in \N$, the solution is given implicitly by
\begin{equation}\label{add_1}
\sqrt{1-bu} F\Big(\frac{1}{2},k,\frac{3}{2};1-bu\Big) = \frac{a}{2b^{k-1}} (x+vt+x_0)
\end{equation}
where
\begin{equation}
a=\sqrt{\frac{4v}{s(\alpha+1)}}, \quad b=\frac{\alpha}{sv},
\end{equation}
and $F$ is the hypergeometric function \cite{lebedev}
\begin{equation}
F(\alpha,\beta,\gamma;z)=\sum_{k=0}^\infty \frac{(\alpha)_k (\beta)_k}{(\gamma)_k} \frac{z^k}{k!}.
\end{equation}
Similarly, for $\alpha=2k$, $k\in \N$, we find
\begin{equation}\label{add_2}
\sqrt{1-bu} F\Big(\frac{1}{2},\frac{1}{2}-k,\frac{3}{2};1-bu\Big) = \frac{a}{2b^{k-\frac{1}{2}}} (x+vt+x_0).
\end{equation}
For $\alpha=1$ (and $s=2$) equation \eqref{add_1} yields the well-known one-soliton solution for
KdV,
\begin{equation}
u(x,t)=2v\, \text{sech}^2 \Big(\sqrt{\frac{v}{4}} (x+vt+x_0)\Big),
\end{equation}
while for $\alpha=2$ it gives a rational solution
\begin{equation}
u(x,t)=\frac{4}{a^2 (x+vt+x_0)^2+4b}.
\end{equation}
The travelling wave solutions for negative integers $\alpha$ are given by similar expressions. For $\alpha=-(2k+1)$, $k\in \N$,
we have
\begin{equation}
\sqrt{1-bu} F\Big(\frac{1}{2},-k,\frac{3}{2};1-bu\Big) = \frac{ab^{k+1}}{2}(x+vt+x_0),
\end{equation}
and for $\alpha=-2k$, $k\in \N\setminus \{1\}$, we find
\begin{equation}
\sqrt{1-bu} F\Big(\frac{1}{2},\frac{1}{2}-k,\frac{3}{2};1-bu\Big) = \frac{ab^{k+\frac{1}{2}}}{2}(x+vt+x_0).
\end{equation}
For $\alpha=-2$ (and $s=1$) we obtain a rational solution of the Harry Dym equation,
\begin{equation}
u(x,t)=(9v)^{\frac{1}{3}} (x+vt+x_0)^{\frac{2}{3}}.
\end{equation}
When $\alpha=-1$ the solution is given as a series $u(x,t)=\sum_{n=0}^\infty c_n (x+vt)^n$. The
coefficients $c_n$ are found recursively from
\begin{equation}
\sum_{k=0}^\infty \alpha_k \beta_{n-k} + \frac{2v}{s}(n-2)c_{n-2}-\frac{2}{s^2}\gamma_{n-2}=0, \quad n\geq 3,
\end{equation}
where $\alpha_n$, $\beta_n$, $\gamma_n$ are defined by
\begin{equation}
\alpha_n = n(n-1)(n-2)c_n, \quad
\beta_n = \sum_{k=0}^n c_{n-k} c_k, \quad
\gamma_n = \sum_{k=0}^n k c_{n-k} c_k.
\end{equation}
The recurrence relation places no restrictions on $c_0$, $c_1$ and $c_2$. For $n\geq 3$ the
coefficient $c_n$ is uniquely determined from $c_0,c_1,\ldots ,c_{n-1}$, hence the solution $u$ is
nontrivial provided $c_0$, $c_1$ and $c_2$ are all nonzero. The existence of travelling wave
solutions is a consequence of the spatio-temporal translational symmetry of equation \eqref{GKdV}.
Although the travelling wave solutions are not directly related to integrability of \eqref{GKdV},
they may prove helpful in investigating the system's properties.

\section{Gauge transformations and the KdV sequences}

In this section we consider more closely the residual gauge transformations for the KdV equation
and their relation with the Lenard recursion operator for the bi-Hamiltonian structure of KdV. We
also investigate gauge transformations between the KdV and related systems, as well as their
residual transformations. If $T=u/2$ ($\alpha=1$, $s=2$) the gauge fields in \eqref{DSfields}
become
\begin{equation}\label{KdVfields}
A_1^{(K)} = \begin{pmatrix} 0 & 1 \\ -\frac{1}{2}u & 0 \end{pmatrix}, \quad
A_2^{(K)} = \begin{pmatrix} -\frac{1}{2}u_x & u \\ -\frac{1}{2}u_{xx}-\frac{1}{2}u^2 & \frac{1}{2}u_x \end{pmatrix},
\end{equation}
hence \eqref{GKdV} yields
\begin{equation}\label{KdV}
u_t = 3u u_x + u_{xxx}.
\end{equation}
This is one of the equivalent forms of the KdV equation. Equation \eqref{symmetry} gives the
residual gauge transformation for the KdV equation,
\begin{equation}\label{KdVsymmetry}
y_t = 3u y_x+y_{xxx}.
\end{equation}
This is the linear equation associated with the KdV equation which Gardner et. al. obtained from the inverse scattering method \cite{gardner}.
If $y$ is a solution of \eqref{KdVsymmetry}, then $\widetilde u = u+ \tau \delta u$ is an infinitesimal transformation of the KdV equation,
where according to \eqref{du2}
\begin{equation}
\delta u (y) = 2u y_x+u_x\, y+y_{xxx}. \label{delta_u2}
\end{equation}
Equation \eqref{KdVsymmetry} has an infinite number of solutions for the parameter function $y$ which are related to a
hierarchy of conservation laws for the KdV equation. Let $y^{(1)}$ be a solution of \eqref{KdVsymmetry} and define the function
\begin{equation}
G^{(1)} \equiv \delta u(y^{(1)}) = 2u\, y^{(1)}_x+u_x\, y^{(1)}+y^{(1)}_{xxx}.
\end{equation}
Then $G^{(1)}$ satisfies the evolution equation
\begin{equation}\label{conservationKdV}
G^{(1)}_t = G^{(1)}_{xxx}+3u_x\, G^{(1)}+3u\, G^{(1)}_x.
\end{equation}
This can be seen by replacing the time derivatives in $G^{(1)}_t$ by \eqref{KdV} and
\eqref{KdVsymmetry}, and expressing $G^{(1)}_t$ only in terms of the $x$-derivatives of $u$ and
$y^{(1)}$. We note that equation \eqref{conservationKdV} can be written as a local conservation law
\begin{equation}\label{FG}
\frac{\partial G^{(1)}}{\partial t}=\frac{\partial F^{(1)}}{\partial x}
\end{equation}
with density $G^{(1)}$ and flux $F^{(1)} = G^{(1)}_{xx}+3u\, G^{(1)}$. Now consider the function
$y^{(2)}$ defined by $y^{(2)}_x = G^{(1)}$. The conservation law \eqref{FG} implies that time
evolution of $y^{(2)}$ is given by
\begin{equation}
y^{(2)}_t = \int G^{(1)}_t dx = 3u\, y^{(2)}_x+y^{(2)}_{xxx},
\end{equation}
which is of the same form as \eqref{KdVsymmetry}. Hence, $y^{(2)}$ also generates a residual gauge
transformation for KdV. By iterating the above procedure we obtain an infinite hierarchy of
residual transformations $y^{(n)}$ defined recursively by
\begin{equation}\label{yG}
y^{(n+1)}_x = G^{(n)}, \quad G^{(n)} = \delta u(y^{(n)}), \quad n\in \N,
\end{equation}
together with local conservation laws
\begin{equation}
\frac{\partial G^{(n)}}{\partial t} = \frac{\partial F^{(n)}}{\partial x}, \quad F^{(n)} = G^{(n)}_{xx}+3u\, G^{(n)}, \quad n\in \N.
\end{equation}
To illustrate the point, we start with the simplest solution $y^{(1)}=1$. Then the first four residual transformations yield
\begin{align}
y^{(1)} &= 1, \\
y^{(2)} &= u, \\
y^{(3)} &= \frac{3}{2} u^2+ u_{xx}, \\
y^{(4)} &= \frac{15}{6} u^3+\frac{5}{2} u_x^2+5 u\, u_{xx}+ u_{xxxx}.
\end{align}
We remark that the functions $y^{(n)}$ are related to the isospectral problem for the Schr\" odinger equation, as
shown in \cite{das}. One can prove that if the eigenvalue $\lambda$ in the Schr\" odinger equation $\Psi_{xx}+\frac{1}{6}u\Psi=
\lambda \Psi$ is $t$-independent, then $u$ must satisfy the evolution equation $u_t=\delta u\left(y^{(n)}\right)$ for some
$n\in \N$. In this way starting from the isospectral problem for the Schr\"odinger operator one can associate to KdV a hierarchy of
integrable equations. A different approach to derive integrable hierarchies using Sato's pseudo differential operator was
given in \cite{kundu}.

Next we show that the functions $y^{(n)}$ are related to an infinite hierarchy of conserved
quantities for the KdV equation as follows. Let $u_t=K[u]$ be an evolution equation for $u(x,t)$
and suppose that the functional
\begin{equation}
\mathcal{H}_n [u]=\int_{-\infty}^\infty H_n[u]\, dx, \quad n\geq 1,
\end{equation}
is a conserved quantity under the flow $u_t=K[u]$. In \cite{wadati} Wadati showed that the
variational derivative of $\mathcal{H}_n$ satisfies the integral equation
\begin{equation}\label{int_eq}
\int_{-\infty}^\infty \left(h(x)\, \frac{\partial}{\partial t}\Big(\frac{\delta \mathcal{H}_n}{\delta u}\Big) + dK(u,h)\,
\frac{\delta \mathcal{H}_n}{\delta u}\right)\, dx =0
\end{equation}
for every function $h(x)$ with compact support, where $dK(u,h)$ is the differential $dK(u,h)=\d0
K(u+\tau h)$. If we take $K[u]=3 u\, u_x+u_{xxx}$, then \eqref{int_eq} implies that $\delta
\mathcal{H}_n/\delta u$ satisfies equation \eqref{KdVsymmetry}. Thus, to each conserved quantity
$\mathcal{H}_n$ of the KdV equation we can associate a residual transformation by
\begin{equation}\label{yH}
y^{(n)} = \frac{\delta \mathcal{H}_n}{\delta u}.
\end{equation}
It is worth noting that there is a close connection between the functions $G^{(n)}$, $y^{(n)}$ and
the Lenard recursion operator for the bi-Hamiltonian structure of KdV \cite{olver}. Indeed, in view
of \eqref{delta_u2} we have $G^{(n)} = Ey^{(n)}$ where $E$ is the differential operator
$E=D^3+2uD+u_x$ and $D$ denotes the derivative with respect to $x$. Then \eqref{yG} implies
\begin{equation}
G^{(n)} = R\, G^{(n-1)},
\end{equation}
where $R=E D^{-1}$ is precisely the Lenard operator for the KdV equation. We also note that
\begin{equation}
y^{(n)}=R^\ast y^{(n-1)}
\end{equation}
where $R^\ast = D^{-1} E$ is the adjoint of $R$. This implies
that the Hamiltonians $\mathcal{H}_n$ defined by \eqref{yH} are related by the Lenard recursion formula
\begin{equation}
D\frac{\delta \mathcal{H}_n}{\delta u} = E\frac{\delta \mathcal{H}_{n-1}}{\delta u}.
\end{equation}
Solving equation \eqref{yH} for $\mathcal{H}_2$ and $\mathcal{H}_3$ we obtain
\begin{equation}
\mathcal{H}_2[u] = \int_{-\infty}^\infty \frac{1}{2} u^2 dx, \quad
\mathcal{H}_3[u] = \int_{-\infty}^\infty \left(-\frac{1}{2} u_x^2+\frac{1}{2}u^3\right)\, dx.
\end{equation}
The Hamiltonians $\mathcal{H}_2$ and $\mathcal{H}_3$ give rise to the well-known bi-Hamiltonian
structure for KdV since
\begin{equation}
u_t = D\frac{\delta \mathcal{H}_3}{\delta u} = E\frac{\delta \mathcal{H}_2}{\delta u}.
\end{equation}
All Hamiltonians $\mathcal{H}_n$ are conserved under the KdV flow, and they are all in involution
with each other with respect to the Gardner-Poisson brackets for the Hamiltonian operators $D$ and
$E$ \cite{lax}. Hence, equation \eqref{yG} can be viewed as another interpretation of the
bi-Hamiltonian property of the KdV equation. Applications of gauge transformations to Hamiltonian
systems from a different point of view can be found in \cite{liu}.

Next we consider the Miura transformation from the gauge theoretical point of view. Miura-type
transformations are known to provide links between different integrable systems, as discussed in
\cite{pelinovsky}. Recall that with the choice $T=u/2$ there is no gauge freedom left for the
matrices $A_1^{(K)}$ and $A_2^{(K)}$ in equation \eqref{KdVfields}, and the zero-curvature
condition is equivalent with the KdV equation \eqref{KdV}. Consider the gauge transformation of
$A_1^{(K)}$ and $A_2^{(K)}$ defined by the matrix
\begin{equation}\label{g}
g=\begin{pmatrix} 1 & 0 \\ -v & 1 \end{pmatrix}
\end{equation}
for some $v=v(x,t)$. The gauge fields transform according to
\begin{align}
A_1^{(M)} &= g\, A_1^{(K)}\, g^{-1}+g_x\, g^{-1}
= \begin{pmatrix} v & 1 \\ -\frac{1}{2}u-v^2-v_x & -v \end{pmatrix},  \\
A_2^{(M)} &= g\, A_2^{(K)}\, g^{-1}+g_t\, g^{-1}
= \begin{pmatrix} u\, v-\frac{1}{2}u_x & u \\ -\frac{1}{2}u^2-\frac{1}{2}u_{xx}-v^2\, u+v\, u_x-v_t & -u\, v + \frac{1}{2}u_x \end{pmatrix}.
\end{align}
Now suppose that $v$ and $u$ are related by the Miura transformation
\begin{equation}
v^2+v_x = -\frac{1}{2}u.
\end{equation}
Then $A_1^{(M)}$ and $A_2^{(M)}$ become
\begin{equation}\label{mKdVfields}
A_1^{(M)} = \begin{pmatrix} v & 1 \\ 0 & -v \end{pmatrix}, \quad
A_2^{(M)} = \begin{pmatrix} -2v^3+v_{xx} & -2(v^2+v_x) \\ -v_t-6v^2\, v_x+v_{xxx} & 2v^3-v_{xx} \end{pmatrix},
\end{equation}
and the zero-curvature condition is equivalent with
\begin{equation}
\left(\frac{\partial}{\partial x}+2v\right) \left(-v_t-6v^2\, v_x + v_{xxx}\right) = 0.
\end{equation}
Since the zero-curvature equation is invariant under gauge transformations,
we recover the well-known result which states that if $v$ satisfies the modified KdV (mKdV) equation
\begin{equation}
v_t = v_{xxx}-6v^2\, v_x,
\end{equation}
then $u=-2(v^2+v_x)$ solves the KdV equation. Note that the reverse implication does not
necessarily hold.

By using the same arguments as in Sec. \ref{general_framework} one can find the residual
transformations for the mKdV equation. Define $K\colon sl(2,\R)\to \R^2$ by
$K\left[\left(\begin{smallmatrix} a & b \\ c & -a \end{smallmatrix}\right)\right]= (b-1,c)$, and
consider the subvariety
\begin{equation}
\mathcal{S}_K =\left\{ A\in sl(2,\R)\mid K(A)=0\right\}
\end{equation}
which consists of matrices of the form $A_1^{(M)}$. Define the gauge group $G=\{\exp(\tau M)\mid \tau \in \R\}$ for some
$M\in sl(2,\R)$. Then
$G$ is a symmetry group of the set $\mathcal{S}_K$ if and only if the infinitesimal gauge transformation by $M$
satisfies
\begin{equation}\label{mKdV_infinitesimal}
\pi \left([M,A]+\frac{\partial M}{\partial x}\right)=0 \quad \text{for all}\quad A\in \mathcal{S}_K,
\end{equation}
where the projection $\pi \colon sl(2,\R)\to \R^2$ is given by $\pi \left[\left(\begin{smallmatrix}
a & b \\ c & -a \end{smallmatrix}\right)\right] =(b,c)$. Evaluation of the condition
\eqref{mKdV_infinitesimal} with $A=A_1^{(M)}$ shows that $M$ can be parameterized by a single
function $z (x,t)$ as
\begin{equation}
M=\begin{pmatrix} z\, v -\frac{1}{2} z_x & z \\ c & -z\, v+\frac{1}{2}z_x \end{pmatrix} \quad \mbox{where}
\quad c=\exp\left(-2\int v\, dx\right),
\end{equation}
Now, the gauge transformation of $A_1^{(M)}$ yields
\begin{align}
\Gamma_{\exp(\tau M)} A_1^{(M)} &= A_1^{(M)}+\tau \left([M,A_1^{(M)}]+\frac{\partial M}{\partial x}\right)+o(\tau^2)  \\
&= \begin{pmatrix} v & 1 \\ 0 & -v\end{pmatrix}+\tau \begin{pmatrix} -\frac{1}{2}z_{xx}+(zv)_x-c & 0 \\
0 & \frac{1}{2}z_{xx}-(zv)_x+c \end{pmatrix}+o(\tau^2),
\end{align}
hence
\begin{equation}
\widetilde v = v+\tau \left(-\frac{1}{2}z_{xx}+(zv)_x-c\right)
\end{equation}
is an infinitesimal transformation of the mKdV equation. This transformation is nonlocal because
$c$ is defined in terms of an indefinite integral. Among the first to consider such transformations
and to point out some applications were Vinogradov and Krasilshchik \cite{vinogradov}. Recently,
nonlocal symmetries for some well-known integrable systems were found in \cite{schiff}.

We are left with the problem of determining the evolution equation for the function $z$. Let
$u=-2(v^2+v_x)$ denote the $(1,2)$ element of the matrix $A_2^{(M)}$ in equation
\eqref{mKdVfields}. The gauge transformation of $A_2^{(M)}$ yields the infinitesimal transformation
of $u$,
\begin{equation}\label{delta_u_eta}
\widetilde u = u+\tau \left(z_t-u z_x+u_x z\right).
\end{equation}
Since $\widetilde u$ and $\widetilde v$ are also related by $\widetilde u = -2({\widetilde v\,}^2+\widetilde v_x)$ to first order in $\tau$,
this implies that $z$ satisfies the equation
\begin{equation}\label{eta}
z_t = z_{xxx}-6 (v^2+v_x)\, z_x.
\end{equation}
We note that this is just the residual symmetry of the KdV equation for the function
$u=-2(v^2+v_x)$. Therefore, if $z$ is a solution of equation \eqref{eta}, then $v\mapsto \widetilde
v = v+\tau \delta v$ is an infinitesimal symmetry of the mKdV equation where
\begin{equation}
\delta v = -\frac{1}{2}z_{xx}+(z\, v)_x-\exp\left(-2\int v\, dx\right).
\end{equation}

The gauge transformation with $g$ of the form \eqref{g} can be repeatedly applied to obtain other integrable systems.
Consider the transformations of $A_1^{(M)}$ and $A_2^{(M)}$ defined by the matrix
\begin{equation}
h=\begin{pmatrix} 1 & 0 \\ -w & 1 \end{pmatrix}.
\end{equation}
We have
\begin{align}
A_1^{(C)} &= h\, A_1^{(M)}\, h^{-1}+h_x\, h^{-1} = \begin{pmatrix} v+w & 1 \\ -2v\, w -w^2-w_x & -(v+w) \end{pmatrix} \\
A_2^{(C)} &= h\, A_2^{(M)}\, h^{-1}+h_t\, h^{-1} \\
&= \begin{pmatrix} -2v^3-2(v^2+v_x)\, w +v_{xx} & -2(v^2+v_x)  \\
2(2v^3-v_{xx})\, w +2(v^2+v_x)\, w^2-w_t & 2v^3+2(v^2+v_x)\, w - v_{xx} \end{pmatrix}.
\end{align}
If we choose $w$ so that the $(2,1)$ element of $A_1^{(C)}$ vanishes, i.e.
\begin{equation}
w_x+w^2+2v w =0,
\end{equation}
then the zero-curvature condition for $A_1^{(C)}$ and $A_2^{(C)}$ yields the Calogero KdV (CKdV) equation \cite{calogero}
\begin{equation}
w_t = w_{xxx}-\frac{1}{2}\left(3 w_x^2 \, w^{-1}+w^3\right)_x.
\end{equation}
Thus, by successively applying the gauge transformations of the form \eqref{g} we obtain a sequence
of integrable equations called the KdV sequence. The successive B\"{a}cklund transformations from
the KdV to CKdV equation were discussed in the bilinear formalism in \cite{nakamura}. The same
sequence of integrable equations was obtained by Pavlov \cite{pavlov} by considering solutions of
the overdetermined system
\begin{equation}
\psi_{xx}=(\lambda-\frac{1}{2}u) \psi, \quad \psi_t = (4\lambda+u)\psi_x-\frac{1}{2}u_x \psi,
\end{equation}
whose consistency condition yields the KdV equation. He showed that by expanding the wave function
into a power series in the spectral parameter $\lambda$,
\begin{equation}
\psi = \exp\Big(\int\Big( \sum_{n=0}^\infty r^{(n)} \lambda^n\Big) dx\Big),
\end{equation}
the coefficients $r^{(n)}(x,t)$ satisfy the same sequence of equations derived above. Thus, the successive transformations
with no spectral parameter in our theory correspond to the expansion of the wave function into power series in
the spectral parameter in conventional theories.

\section{The Harry Dym Equation}

The Harry Dym (HD) equation \cite{kruskal}
\begin{equation}\label{HD}
u_t = -\frac{1}{4} u^3 u_{xxx}
\end{equation}
is obtained from \eqref{GKdV} by setting $\alpha=-2$ and $s=1$. It was discovered by Harry Dym when
trying to transfer some results about the isospectral theory to the string equation \cite{hereman}.
The HD equation is a completely integrable Hamiltonian system which is solvable by the inverse
scattering transform \cite{wadati2}, \cite{wadati3}, and it possesis the bi-Hamiltonian structure
and infinitely many conservation laws \cite{brunelli}.

In this case the fields in \eqref{DSfields} become
\begin{equation}
A_1^{(HD)} = \begin{pmatrix} 0 & 1 \\ -\frac{1}{u^2} & 0 \end{pmatrix}, \quad A_2^{(HD)} = \begin{pmatrix} -\frac{1}{2} u_x & u \\
-\frac{1}{2} u_{xx}-\frac{1}{u} & \frac{1}{2}u_x \end{pmatrix}.
\end{equation}
In view of equations \eqref{symmetry} and \eqref{du2} the infinitesimal symmetry transformation for
the HD equation is given by
\begin{equation}
\widetilde u = u+\tau (-u\, y_x+u_x\, y-\frac{1}{4}u^3 y_{xxx}),
\end{equation}
where $y$ is constrained by the equation
\begin{equation}\label{HDresidual}
y_t = -\frac{1}{4} u^3 y_{xxx}.
\end{equation}
We show that equation \eqref{HDresidual} has an infinite number of solutions for the parameter
function $y$. Suppose $y^{(1)}$ is a solution of \eqref{HDresidual} and define $y^{(2)}$ by
\begin{equation}\label{y2}
\left(\frac{y^{(2)}}{u}\right)_x = - u\, y^{(1)}_{xxx}.
\end{equation}
We claim that $y^{(2)}$ is also a solution of equation \eqref{HDresidual}. Denote for the moment
$y^{(1)}_{xxx}=v$ so that $y^{(2)}=-u \int uv\, dx$. Then a simple computation shows that
\begin{equation}
y^{(2)}_t = \frac{1}{4} u^3 u_{xxx} \int uv\, dx+\frac{1}{4} u \left(\int u_{xxx} u^3 v\, dx + \int u (u^3 v)_{xxx}\, dx\right)
\end{equation}
where we used the assumption that $y^{(1)}$ satisfies \eqref{HDresidual}, and we replaced the time
derivative of $u$ by equation \eqref{HD}. After a few partial integrations of the last term we
obtain
\begin{equation}
\int u (u^3 v)_{xxx}\, dx = u (u^3 v)_{xx}-u_x (u^3 v)_x+u_{xx} u^3 v - \int u_{xxx} u^3 v\, dx,
\end{equation}
which finally yields
\begin{equation}
y_t^{(2)} = \frac{1}{4} u^3 \left(u_{xxx} \int uv\, dx+3 u_x^2 v+5 u u_x v_x + 4 u u_{xx} v + u^2 v_{xx}\right).
\end{equation}
It is straightforward to verify that the expression in the parentheses equals $-y^{(2)}_{xxx}$,
hence $y^{(2)}$ satisfies equation \eqref{HDresidual}. Therefore, \eqref{y2} provides a recursion
formula for $y^{(n)}$ which generates an infinite hierarchy of residual gauge transformations for
the HD equation. Starting with $y^{(1)}=1$ the first four functions are given by
\begin{align}
y^{(1)} &= 1, \\
y^{(2)} &= u, \\
y^{(3)} &= \frac{1}{2} u\, u_x^2-u^2 u_{xx}, \\
y^{(4)} &= \frac{3}{8} u\, u_x^4-\frac{3}{2} u^2 u_x^2 u_{xx}+\frac{3}{2} u^3 u_{xx}^2 +2 u^3 u_x\, u_{xxx}+u^4 u_{xxxx}.
\end{align}
As in the case of the KdV equation the conserved quantities of the HD equation are related to the
residual gauge transformations $y^{(n)}$. Let
\begin{equation}
\mathcal{G}_n[u]=\int_{-\infty}^\infty G_n[u]\, dx, \quad n\in \N,
\end{equation}
be a conserved quantity of the HD equation \eqref{HD}. Substituting $K[u]=-\frac{1}{4}u^3 u_{xxx}$
into equation \eqref{int_eq} one can show that $\delta \mathcal{G}_n/\delta u$ satisfies
\begin{equation}\label{Gn}
\left(\frac{\delta \mathcal{G}_n}{\delta u}\right)_t = \frac{3}{4} u^2 u_{xxx}\, \frac{\delta \mathcal{G}_n}{\delta u}
-\frac{1}{4} \left(u^3 \frac{\delta \mathcal{G}_n}{\delta u}\right)_{xxx}.
\end{equation}
We emphasize that, unlike in the KdV case, the above equation is not the same as the residual
transformation \eqref{HDresidual} for $y^{(n)}$. However, if we set
\begin{equation}\label{yGn}
y^{(n)}=u^3\, \frac{\delta \mathcal{G}_n}{\delta u},
\end{equation}
then it is easily seen that $y^{(n)}$ satisfies \eqref{HDresidual} if and only if $\delta
\mathcal{G}_n/\delta u$ satisfies \eqref{Gn}. Therefore, the conserved quantities and residual
gauge transformations of the HD equation are related by equation \eqref{yGn}. The conserved
quantities corresponding to the first four residual transformations $y^{(n)}$ found earlier are
given by
\begin{align}
\mathcal{G}_1[u] &=\int_{-\infty}^\infty \Big(-\frac{1}{2}u^{-2}\Big)\, dx, \\
\mathcal{G}_2[u] &=\int_{-\infty}^\infty \left(-u^{-1}\right)\, dx, \\
\mathcal{G}_3[u] &=\int_{-\infty}^\infty \frac{1}{2}u^{-1} u_x^2\, dx, \\
\mathcal{G}_4[u] &=\int_{-\infty}^\infty \left(\frac{1}{8}u^{-1} u_x^4+\frac{1}{2}u u_{xx}^2\right)\, dx.
\end{align}

\section{The sine-Gordon and nonlinear Schr\"{o}dinger equations}

So far we have discussed integrable systems on the Lie algebra $sl(2,\R)$ which arise by fixing the
DS gauge \eqref{DS}. In this section we consider briefly a few generalizations obtained by changing
the gauge fixing condition and generalizing the gauge group. More precisely, we derive the
sine-Gordon (SG) and focusing nonlinear Schr\"{o}dinger (NLS) equations. The NLS equation was
derived by symmetry reduction from the self-dual Yang-Mills equations in \cite{mason} (see also
\cite{ablowitz}). We remark that the residual transformations approach presented in thet previous
sections can be extended to study infinitesimal symmetries of the SG and NLS equations as well.

Consider the gauge fields \eqref{fields} with $R=\alpha$, $\alpha \in \R\setminus \{0\}$, and $S=T=-\frac{1}{2}w_x$ for some
$w=w(x,t)$. In order to fix the remaining gauge freedom suppose $q=u$. Then the zero-curvature equations \eqref{ZCC1}-\eqref{ZCC3}
yield
\begin{align}
p_x &= -q\, w_x, \label{SG1} \\
\frac{1}{2} w_{xt}+q_x &= p\, w_x + 2\alpha\, q, \label{SG2} \\
-\frac{1}{2} w_{xt}+q_x &= p\, w_x-2\alpha\, q. \label{SG3}
\end{align}
Adding and subtracting equations \eqref{SG2} and \eqref{SG3} we obtain
\begin{align}
q_x &= p\, w_x, \label{SG4} \\
w_{xt} &= 4\alpha\, q. \label{SG5}
\end{align}
We note that equations \eqref{SG1} and \eqref{SG4} admit solutions $p=\cos(w)$ and $q=\sin(w)$.
With this choice of $p$ and $q$ the gauge fields become
\begin{equation}
A_1^{(SG)} = \begin{pmatrix} \alpha & -\frac{1}{2} w_x \\ \frac{1}{2} w_x & -\alpha \end{pmatrix}, \quad
A_2^{(SG)} = \begin{pmatrix} \cos(w) & \sin(w) \\ \sin(w) & -\cos(w)\end{pmatrix}.
\end{equation}
The zero-curvature condition is then equivalent with \eqref{SG5}, which is the sine-Gordon equation
\begin{equation}
w_{xt} = 4\alpha\, \sin(w).
\end{equation}

Similarly, we derive the focusing NLS equation by considering the gauge group $SL(2,\C)$ and introducing the gauge
fixing conditions $R=0$, $S=\overline{T}=w$ and $q=-\overline{u}$. In this case the zero-curvature equations
\eqref{ZCC1}-\eqref{ZCC3} yield
\begin{align}
-q\, w+\overline{q}\, \overline{w} &= -p_x , \label{NLS1} \\
w_t + \overline{q}_x &= 2p\, w, \label{NLS2} \\
\overline{w}_t + q_x &= -2p\, \overline{w}. \label{NLS3}
\end{align}
Equation \eqref{NLS1} implies that $p_x$ is purely imaginary, hence we may impose further reduction
of the gauge freedom by requiring $p=-i |w|^2$. Then equations \eqref{NLS2} and \eqref{NLS3} are
complex conjugate of each other. Furthermore, if we set $q=-i \overline{w}_x$, then relation
\eqref{NLS1} holds identically. In this case the gauge fields are
\begin{equation}
A_1^{(NLS)} = \begin{pmatrix} 0 & w \\ -\overline{w} & 0 \end{pmatrix}, \quad
A_2^{(NLS)} = \begin{pmatrix} -i |w|^2 & -i w_x \\ -i \overline{w}_x & i |w|^2 \end{pmatrix},
\end{equation}
and \eqref{NLS3} yields the focusing NLS equation
\begin{equation}
iw_t - w_{xx}-2 |w|^2\, w = 0.
\end{equation}
The sine-Gordon and the focusing NLS equation can also be derived within the framework of Birkhoff factorization of loop groups.
For more details see e.g. \cite{kresic} and \cite{dorfmeister2}.

\section{Conclusion}

In this paper we have discussed integrable systems within the framework of gauge theory. The gauge
theoretical structure of the Zakharov-Shabat formalism has been fully discussed. Starting with the
DS gauge, we have fixed the residual gauge freedom by the relation \eqref{condition3} and obtained
a two-parameter family of nonlinear equations \eqref{GKdV}. It was shown explicitly that for
$\alpha=1$ (KdV) and $\alpha=-2$ (HD) there exist infinite hierarchies of residual gauge
transformations. The transformations can be calculated recursively and they lead to an infinite
number of conserved quantities. We have also shown that all equations \eqref{GKdV} admit travelling
wave solutions which are given implicitly in terms of the hypergeometric function, except for
$\alpha=\pm 1, \pm 2$ when the solutions are given in explicit form. However, it is an open problem
to determine if equation \eqref{GKdV} is integrable for all $\alpha\in \Z\setminus \{0\}$. The
Miura transformations which lead to the KdV sequence (KdV $\to$ mKdV $\to$ CKdV, etc.) have been
formulated in terms of gauge transformations acting on the gauge fields defined on the Lie algebra
$sl(2,\R)$. It was shown that other types of 1+1 integrable systems can also be formulated within
the same framework by changing the initial gauge fixing condition and generalizing the gauge group.


\end{document}